\documentclass[manuscript,nonacm]{arxiv}

\usepackage{pifont}
\usepackage{multirow}
\usepackage{placeins}

\usepackage{amsmath}
\usepackage{multirow}
\usepackage{xcolor}
\usepackage{array}
\usepackage[shortlabels]{enumitem}
\usepackage[flushleft]{threeparttable}
\usepackage{hyperref}
\usepackage{afterpage,lscape}
\usepackage{csquotes}
\usepackage{pdfpages}
\usepackage{cleveref}

\newcommand{\beginsupplement}{%
        \setcounter{table}{0}
        \renewcommand{\thetable}{S\arabic{table}}%
        \setcounter{figure}{0}
        \renewcommand{\thefigure}{S\arabic{figure}}%
     }

\setcopyright{rightsretained}
\copyrightyear{2024}

\begin{document}

\title{Reimagining AI in Social Work: Practitioner Perspectives on Incorporating Technology in their Practice}


\author{Katie Wassall}
 \affiliation{%
   \institution{University of Cambridge}
   \city{Cambridge}
   \country{UK}}

\author{Carolyn Ashurst}
 \affiliation{%
  \institution{The Alan Turing Institute}
  \city{London}
  \country{UK}
}

\author{Jiri Hron}
 \affiliation{%
   \institution{University of Cambridge}
   \city{Cambridge}
   \country{UK}
}

\author{Miri Zilka}
 \affiliation{%
   \institution{University of Cambridge}
   \city{Cambridge}
   \country{UK}}
 \email{mz477@cam.ac.uk}

\renewcommand{\shortauthors}{Wassall et al.}


\begin{abstract} \looseness=-1
    There has been  a surge in the number and type of AI tools being tested and deployed within both national and local government in the UK, including within the social care sector. 
    Given the many ongoing and planned future developments, the time is ripe to review and reflect on the state of AI in social care. 
    We do so by conducting semi-structured interviews with UK-based social work professionals about their experiences and opinions of past and current AI systems.
    Our aim is to understand what systems would practitioners like to see developed and how.
    We find that all our interviewees had overwhelmingly negative past experiences of technology in social care, unanimous aversion to algorithmic decision systems in particular, but also strong interest in AI applications that could allow them to spend less time on administrative tasks.
    In response to our findings, we offer a series of concrete recommendations, which include commitment to participatory design, as well as the necessity of regaining practitioner trust.
\end{abstract}

\maketitle

\vspace{-0.5\baselineskip}
\section{Introduction}

\looseness=-1
Artificial Intelligence (AI) is increasingly used to inform decision-making in consequential settings within criminal justice, welfare, and social work \citep{ZilSarWel2022, kawakami2022care, johnson2004effectiveness, saxena2020human}. 
In social work, AI-based technologies are being incorporated across the board, including into practitioner decision-making \citep{jorgensen2022three,kawakami2022improving,dencik2018data,englandAdultSocial}.
Adoption is driven by increased demand for efficiency stemming from 
unprecedented pressures on budgets, rising citizen demand, and challenges in the recruitment and retention of workers \citep{devlieghere2017logic}.
At the same time, the social work profession has been shaken by a number of high publicity failures, resulting in a greater 
focus on risk reduction,
and the introduction 
of more data-based assessments \citep{devlieghere2022dataism}. 

\looseness=-1
A common response is the deployment of \emph{algorithmic decision systems} (ADS), under the promise of increased objectivity, efficiency, and accuracy.
ADS are often promoted as `evidence-based', and
as a route to better care without additional human resources.
%
However, ADS and other technological deployments have often failed to deliver the advertised benefits, and instead introduced new risks and harms.
For example, \citet{kawakami2022improving} show that practitioners can find ADS obstructive rather than helpful, and treat the tool as an adversary.
\citet{cheng2022child} then demonstrate that the recommendations of a child welfare screening algorithm show significant racial disparities compared to human decision-makers.

\looseness=-1
In the UK, both the national government and local councils have seen a surge in the number and type of AI tools being tested and deployed \citep{nhsx,dencik2018data}. 
With many ongoing and planned future developments, the time is ripe to review and reflect on the state of AI in social care.
While most prior research focuses on the impacts of specific systems, we take a broader view, aiming to explore practitioners' experiences, perspectives, and needs.
Our goal is to inform future developments both in terms of \textit{what} systems should be developed and \textit{how}.

We interview social workers in the UK to understand their perspectives on AI in social work, with focus on their professional experience with technology, their hopes and concerns for future, and their perspective on participation and design desiderata.
Analysis of the participant responses revealed three key topics:
\begin{enumerate}[(i),topsep=0.5pt]
    \item \looseness=-1
    Overwhelmingly
    negative past experiences of technological systems incorporated into social work. 

    \item \looseness=-1
    Unanimous aversion to the use of ADS within social work.

    \item \looseness=-1
    An interest in non-ADS AI applications, conditioned on these being developed in a participatory manner. 
\end{enumerate}
Our findings suggest a disconnect between organisational and institutional pursuits, and the views and attitudes of practitioners.
Taking a practitioner led approach, we discuss novel design opportunities and future AI applications envisioned by our interviewees.

\looseness=-1
The paper structure is as follows: \Cref{sec:background} familiarises the reader with the sector's emerging use of AI and related works;
\Cref{sec:methodology} describes the methodology;
\Cref{sec:experience,sec:potential_applications,sec:future_development} report on the core themes and findings from our practitioner interviews;
\Cref{sec:discussion} discusses key themes and design visions, and provides a summary of considerations for future developments; \Cref{sec:conclusion} concludes the paper.

\section{Background and related work}
\label{sec:background}

\subsection{Social Work in the UK}

\looseness=-1
Social workers deliver care and support to many people including the elderly, children with disabilities, children at risk of neglect or abuse, young offenders, teenagers with mental health problems, adults with learning disabilities, mental or physical disabilities, people suffering from substance misuse, refugees and asylum seekers, foster carers, and adopters. 
Very broadly, the profession can be broken down into two primary areas: (i)~adult social care, and (ii)~children and families. 
In 2021, Social Work England had 99,191 registered social workers \citep{england2022social}. The majority are employed by social care departments run by local authorities but some social workers are employed by other public bodies (e.g. National Health Service (NHS) or the Children and Family Court Advisory and Support Service), or the voluntary and private sector.

\subsection{AI in social work}

\looseness=-1
Due to unprecedented pressures on budgets, ageing populations, and a shift towards evidence-based decision making, there is growing interest in the potential of AI to support decision making in social care settings. However, there is currently limited evidence of the systems' effectiveness \citep{cresswell2020investigating,clayton2020machine}. 
While a comprehensive review of AI use in social care is out of scope, we provide illustrative examples that typify popular uses and demonstrate common challenges.

\subsubsection{Children and families social care.}
Over the past two decades, multiple US child welfare agencies have begun incorporating ADS into 
the child protection decision making process \citep{brown2019toward,johnson2004effectiveness,kawakami2022improving,saxena2020human}. Two well-documented currently used ADS are the proprietary Rapid Safety Feedback program, and the Allegheny Family Screening Tool (AFST) 
\citep{parker2022examining,vaithianathan2017developing,glaberson2019coding}. 

\looseness=-1
AFST
provides hotline screeners with a risk score to help them decide whether to open an investigation of child abuse or neglect. 
Pulling together data from a number of county systems that document family relationships with public services (including jail, juvenile prison, public welfare, health and census systems), the tool relies on over 100 variables to calculate the final score.
Although some see AFST as a more accurate and transparent way to screen cases \citep{drake2020practical}, there are concerns it unfairly targets children from poor and minority families \citep{glaberson2019coding,chouldechova2018case}. 
Notably, several similar tools have been discontinued because practitioners found them unreliable or opaque \citep{glaberson2019coding,brown2019toward}. 

In the UK, adoption of ADS and predictive analytics is within the discretion of local authorities.
In \citeyear{dencik2018data} England, 53 out of 152 local authorities were using predictive tools \citep{dencik2018data,dencik2019golden}.
This includes tools deployed in high-risk use cases such as identifying children at risk of abuse or neglect \citep{dencik2018data,mcintyre2018councils,redden2020datafied}. 
Following public backlash, Hackney Council halted its use of a commercially developed ADS stating that it did not yield the expected benefits \citep{turner2019using}.

\vspace{-0.5\baselineskip}
\subsubsection{Adult social work}
\looseness=-1
ADS is not the only type of a data-driven system increasingly deployed in adult social care.
\citet{morgan2021artificial} lays out several proposals,
ranging from monitoring, alerts, and task support, to AI companionship. 
The latter has been adopted in Japan, with government spending of over \$300 million on research and development \citep{technologyreviewInsideJapans}. 
However, there is evidence the robots meant to assist caregivers within care homes create more issues than solve \citep{technologyreviewInsideJapans}. 
Another category of innovation is monitoring and `desirable surveillance' of older adults who wish to maintain independence, such as human activity recognition and automatic monitoring \citep{park2016depth}.
While some work well in lab settings---e.g., \citet{park2016depth} claim they 99.55\% average recognition accuracy---limited data is available on real-world use cases.
In the UK, the NHSX AI Lab \citep{nhsx}
has been piloting multiple use cases, including remote monitoring, predictive analytics, and assessing pain in people with dementia \citep{englandAdultSocial}.

\vspace{-0.5\baselineskip}
\subsection{Related work}
 

\looseness=-1
In the wider public sector context,
several studies have investigated on practitioner perspectives on the use of AI, in particular ADS \citep{veale2018fairness,brown2019toward,kawakami2022improving}.
A common thread is the general disregard for the context or the users in development of ADS.
Public sector practitioners have viewed existing ADS deployments as unsuccessful, finding the tools to be a burdensome rather than helpful
\citep{brown2019toward,kawakami2022improving}.
The authors highlight the need to re-think and re-design the \textit{``interfaces, models, and organizational processes that shape the ways ADS tools are used in child welfare, and other public sector decision making contexts''}.
While relevant, the above works differ from ours in their focus on retrospective analysis 
(i.e., post system deployment), 
and focus purely on ADS tools, rather than AI more generally.


\looseness=-1
Several related works did take a forward-looking approach. 
\citet{kawakami2022care} interviewed social workers and co-generated `design concepts', envisioning ways to redesign ADS interface.
\citet{stapleton2022imagining} ran workshops with various stakeholders---predominantly impacted parents and caseworkers---around the design and use of predictive risk models (PRMs) in US child welfare.
The paper explores both concerns regarding the existing systems as well as future avenues of working in collaboration with impacted communities.
\citet{Wang2023we} asked migrant jobseeker helpers to create `design fictions' of objects from 2050 they would use in their work. 
In contrast to the above work, our focus is on the UK, and on understanding how the practitioners' needs, attitudes, and past experiences can shape the next generation of AI-based applications in social care, i.e., not limited to ADS or PRMs.

\section{Methodology}
\label{sec:methodology}

\looseness=-1
We perform a qualitative study of UK social worker perspectives on AI, using a participatory practitioner-focused approach.\footnote{`Practitioner' refers to anyone who does or did practise social care, regardless of their experience and use of AI.}
In February to March 2023, we conducted thirteen individual hour-long semi-structured interviews aiming to both understand practitioner attitudes and experiences with AI, and to identify places where they felt AI can (not) bring value in the future.

\subsection{Semi-structured interviews}

\paragraph{Design.}
\looseness=-1
Following \citet{schultze2011designing}, we used a semi-structured interview format,
as it allows for
an in-depth exploration of individuals’ personal stories, reflections, and reasoning, whilst maintaining a broadly consistent line of questioning \citep{buck2022general}.
The questions were divided into three sections:
\begin{enumerate}[(a),topsep=0.5pt]
    \item 
    current situation and challenges,

    \item 
    future opportunities and concerns,

    \item 
    design and participation.
\end{enumerate}

\looseness=-1
Before the interview, each participant was provided with an information sheet\footnote{For the purposes of review, the sheet has been redacted to preserve authors' anonymity.} (\Cref{apx:info_sheet}) and interview questions (\Cref{apx:guide}), so that they could understand the purpose of our project, and give informed consent.
Both documents (\Cref{apx:info_sheet,apx:guide}) were reviewed and approved by the Head of Research and Policy at the \emph{British Association of Social Workers} (BASW). 
We made all interviewees aware their participation is voluntary, and that responses will remain anonymous.
All interviews were conducted online over Zoom.
Our research received ethical approval from our institutional Research and Ethics Committee.
`AI' has been used as an umbrella term throughout the discussions.
Towards the beginning of each interview, we would prompt a discussion around what the participant understood by the term AI and provide guidance where required. 
This ensured that our discussions were grounded in the same broad understanding of the term.

\paragraph{Participants}

\looseness=-1
Participant recruitment 
was facilitated by BASW, an organisation representing over 22,000 UK social work professionals.
Participants were recruited in two ways.
First, the BASW head of policy and research personally reached out to potential participants based on their sector knowledge, or frontline work experience;
nine of sixteen chose to partake.
Second, an advertisement was shared with all frontline workers via an internal BASW bulletin;
this yielded four additional participants.
The final thirteen recruited participants cover a wide range of specialisations and roles, including frontline social workers from the statutory sector in children and adult services, individuals working in policy or senior management, and qualified social workers currently full-time in academia.
Frontline practitioners spanned both general and specialist areas.
See \Cref{tab:participants} for aggregated participant information.\footnote{\looseness=-1
To try to avoid the common pitfall where participatory research yields no benefit to the participants who sacrificed their valuable time, our high-level findings and recommendations were shared with everyone involved via a short report.}

\looseness=-1
Experience with AI varied among the participants.
We believe it is important to hear the voices of participants with limited experience of AI but deep knowledge of the sector, though we acknowledge this is not a universal view.
The interviewers provided adequate support where needed (e.g., by clarifying vocabulary and capabilities of existing AI tools).

\begin{table}[htb]
    \caption{Aggregated participant demographics}
    \centering
    \begin{tabular}{llr}
    \textbf{Information type} & \multicolumn{2}{l}{\textbf{Participant counts}}  \\  
    \hline
    Research participants & \multicolumn{2}{r}{13} \\
    \midrule
    Qualified social workers  & \multicolumn{2}{r}{10}   \\ 
    \midrule
    \multirow{3}{*}{Stakeholder group}  & Academic & 3 \\ 
                                        & Policy & 5 \\ 
                                        & Frontline & 5  \\
    \midrule
    \multirow{3}{*}{Areas of social work}   & Children & 4 \\
                                            & Adult & 4 \\
                                            & Both & 5 \\
    \midrule
    \multirow{5}{*}{Practice in the field (years)}  
    &  $<5$ & 3 \\ 
    & $5-9$ & 3 \\ 
    & $10-15$ & 2  \\ 
    & $>15$ & 2 \\
    & undisclosed & 4 \\ 
    \midrule
    \multirow{2}{*}{Gender} 
    & Male & 8 \\
    & Female & 5 \\
    \end{tabular} 
    \label{tab:participants}
    \vspace{-1em}
\end{table}

\paragraph{Data collection and analysis}
\looseness=-1
All data capture and analysis were conducted independently of BASW.
All interviews were recorded and transcribed using \texttt{otter.ai}.\footnote{\texttt{otter.ai} is paid service for automated transcription. See \url{https://otter.ai/}.}  
The transcripts were proofread and annotated 
to sort key findings into higher-level themes.
During analysis, participants were grouped into three categories: 
1)~policy;
2)~academia;
3)~frontline. 
All quotes presented in the following sections have been anonymised;
throughout, we make clear if a conclusion was drawn based on general trends, or a response of one or few individuals.

\subsection{Limitations}

\looseness=-1
Firstly, the participants were either recruited through the professional network of the BASW head of policy and research,
or self-selected by replying to the bulletin advert.
Our findings thus may not fully represent all voices across the sector. 

\looseness=-1
Secondly, we are ultimately guided by our experiences and positionality, which are not grounded in professional or personal experience of the social work practice.
The resulting design visions were co-created during the interviews. 
While some interviewees came up with fully fleshed out ideas without our input, several interviewees asked questions about what is feasible from a technological perspective, influencing their vision.
We acknowledge that a different approach to ideation, e.g., if we asked the practitioners to complete the design visions alone (without guidance on technical limitations), or involved several practitioners at a time, might have resulted in different visions. 

\looseness=-1
Finally, we focus only on the views of social workers. 
As discussed \Cref{sec:key_find_lit}, it is necessary that perspectives of all parties impacted by the tools---particularly the carees---are taken into account in future developments.

\section{Theme 1: Experience with technology in social work}
\label{sec:experience}

\looseness=-1
In this and the next two sections, we summarise the major themes that materialised in our interviews. 
Whilst general attitudes towards AI varied, an overall broad consensus emerged around
(i)~negative past experience of technological systems, 
(ii)~unanimous aversion to ADS, and
(iii)~the importance of practitioner-centred design. 
These key findings emerge in Sections \ref{sec:experience}, \ref{sec:potential_applications}, and \ref{sec:future_development} respectively.

\subsection{Awareness of AI in social work}

\subsubsection{Levels of exposure to AI} 
\looseness=-1
Among the thirteen interviewees, there was a wide spectrum of understanding and experience of AI in social work practice. Whilst all participants were familiar with the term ‘AI’, specific understanding of the term and experience of it varied among the different stakeholder groups (academics, policy, and frontline). Unsurprisingly, the academics working in this field were the most knowledgeable and forthcoming with different examples of AI in social care and social work practice, followed closely by policymakers, and then finally the practitioners themselves who appeared to have little exposure to AI in their day-to-day practice. 

\subsubsection{Examples of AI in social work}
\looseness=-1
Both the academic and policy participants shared a number of first-hand examples of their experience of AI.
For example, one participant discussed \textit{Tribe}, a smart technology company which aims to intelligently map, develop, and connect people seeking care, with care providers.
They characterised Tribe as 
\textit{``trying to map areas of the country where there are \ldots 
very few formal services available \ldots to help visualise them so you can meet those needs''}. 
Other examples included \textit{Argenti} (a telecare provider of assisted living technologies),\footnote{Assistive technology refers to a large range of devices or systems that help people maintain or improve a carrying out tasks related to everyday life. Only some of these, e.g., intelligent monitoring, are AI-based.} \textit{Brain in Hand} (a digital self-management support system for people who need help remembering things, making decisions, planning, or managing anxiety), \textit{PredictX} (a predictive analytics platform), and \textit{Pepper} (a robot designed to assist people). 

\looseness=-1
It quickly became apparent there are \textit{``a lot of examples of AI being used in the independent care sector''} (P2).
A particularly interesting one was shared by P4 who currently works on evaluating a grassroots developed technology for child protection social workers. 
The application seeks to overcome the modularised siloed nature of the case management process, by creating a shared space that can be used collaboratively by the social worker and families alike.
The aim is to open up communication, increase transparency, and empower families.

\looseness=-1
In contrast, four of the five practising social workers had little exposure to AI in their day-to-day work.
One of the practitioners said they decided to participate due to their perception that \textit{``AI in social work is an area that I never thought would get any sort of looking into''} (P8).
Another practitioner highlighted \textit{``I've never seen an obvious example of a process that's got AI designed into it''} (P3). 
One frontline practitioner who did have first-hand experience of AI in social work said \textit{``people have tried to use it for risk management, problem families, problem children''}, but also recognised \textit{``I probably have more of an interest and more general awareness than most social workers''} (P7).

\subsection{Perceived benefits and risks}

\subsubsection{Potential benefits}
In spite of the lack of exposure to AI in their day-to-day roles, there was a clear interest from all frontline practitioners around the its potential. 
For example, \textit{``something in the new generation of tech that could transform the way that social care is offered, if it's done properly, if it's done ethically''} (P3).
This sentiment was echoed by one of the policy makers: 
\textit{``[S]ome aspects, I think, that can be supported and enhanced through the use of AI, particularly in relation to needs that relate to people's activities of daily living, around physical disabilities, but also potentially, in relation to the need for connection with other people, for information, or for advice''} (P6).

\looseness=-1
A particularly interesting view was expressed by a participant who recently moved from frontline practice to pursuing a PhD in social work:
\textit{``My first reaction is that I don't think it's a good idea \ldots
It goes against social values and the kind of social work skills which are inherent to the profession.
\ldots
[However,]
I suppose since coming away from social work and doing research, I'm a bit more open to using technology for good and more effective children's services, job protecting systems, and things, \ldots but it has to be limited to very specific uses''} (P4).

\subsubsection{Potential risks}

\looseness=-1
This concern around AI and the need for a deeper understanding of the risks was echoed by a several participants:
\textit{
``[I]f we're not digitally critical of AI, it could run away with us. 
\ldots
[I]t's like having a football match without a football pitch. 
If you don't have the boundaries, it just 
goes wherever it wants to'' (P2). 
}
P10 highlighted:
\textit{``There is a growing realisation that there's still quite a lot of limitations and risks involved in the use of different technological applications.
We're still in a very early stage, both in terms of their development, but also in terms of a broader understanding of them''}.

\subsection{Lack of alignment between stakeholders}

\subsubsection{Overpromised and underdelivered}
\looseness=-1
In spite of participants' interest in the potential of AI in social work, there was shared scepticism of the decisions and rhetoric of senior management. 
For example, P1 stated: ``\textit{my instinctive reaction is to be suspicious, partly because my experience over my career is being overpromised and then feeling underdelivered to in terms of the benefits of technology''}. 
P9 echoed this sentiment: 
\textit{``There's 
this magical belief that technology will improve productivity in the public sector and in social care''}. 

\looseness=-1
Discussions revealed that for many of the participants, this suspicion can largely be attributed to a breakdown in trust: 
\textit{``I think there is a real trust issue.
Frontline workers feel they've been sold technology as a solution by the managers before and it hasn't worked''} (P12). 
The mistrust has developed from past experiences with deployments of technology, and concerns around \textit{``who's steering the conversation? No one is proactively developing best practices''} (P2). 
P7 considers the sectors move toward digitisation as a fundamental challenge to their role: 
\textit{``The way it's been introduced coincided with a move away from a more social approach to social work, to a more \ldots risk management approach, and the technology was constructed and designed to do that, 
rather than help social workers do social work''}.

\vspace{-0.25\baselineskip}
\subsubsection{Commercial interests contributing to overpromising}
\looseness=-1
Several participants commented that commercial organisations selling technological solutions contribute to the overhyped claims. 
One participant shared first-hand experience in relation to an attempt of the UK government to understand the specific age of unaccompanied minor refugees:   
\textit{``I'm sure that there are people trying to find a way of monetizing this, and convincing the government \ldots they've got some neural network that can get more out of the MRI scans than a human being can''} (P7).
When discussing the lack of trust, P12 went on to note that \textit{``the public is suspicious of big tech, and the use to which data will be put''}.

\vspace{-0.25\baselineskip}
\subsubsection{Lack of alignment between developers and practitioners}
\looseness=-1
All participants felt strongly that those who commission and develop technological solutions lack understanding of practitioners' context,
which
contributed to many of the past technological failings. 
For example:  
\textit{
``[T]he tech company couldn't imagine a world in which an office didn't have a computer. 
And a care worker couldn't imagine a world in which there was one''} (P1). 
\textit{``What's been really interesting is seeing this tension between people who have a social work background and the app developers''} (P4). 

\vspace{-0.25\baselineskip}
\subsection{Prior technology failures in social work}

\subsubsection{Poor user experience}
\looseness=-1
Participants described how the lack of developer understanding of the practitioners' context
resulted in technological interventions that cater to the needs of neither the practitioners, nor the carees.
There was unanimous agreement that the majority of the existing digitised systems---broadly, not just AI---are \textit{``really awful to use, \ldots rigid, and burdensome''} (P4).
Instead of helping practitioners, existing systems are broadly seen as a barrier: \textit{``[T]o give people the best chance of the care that they needed, the most person-centred care, was just consistently a pain in the backside. It didn't matter where I was geographically''} (P2).

\vspace{-0.25\baselineskip}
\subsubsection{Lack of practitioner involvement} 
\looseness=-1
The lack of practitioner involvement was quoted as a major barrier to positivity around new technology, including AI.
Interviewees mostly attributed past technological failures to the design without practitioner participation. 
The feeling that \textit{``we are not adequately involved in the decision making process or in the implementation process''} (P1) led to an even greater disengagement: \textit{``the opinions of the practitioner will never be heard when these tools developed''} (P9). 
The following example was shared by P10:
\textit{``[A]n ICT system was brought in around 2010.
The government decided that 
\ldots
child protection 
\ldots
needed 
a
uniform standardised approach. 
No social workers were engaged in developing the system, so it didn't look at what social workers needed. 
\ldots
And guess what? The system flunked because it wasn't addressing the issue that was actually there, and it hadn't had input from the people who were dealing with it on the front line.''}

\looseness=-1
One participant did have experience of being consulted, but \textit{``not much of it made it into the final product''} (P13). 
P2 summarised: 
\textit{``For social work, it's the blind leading the blind. 
We don't have system leaders
\ldots
[or] social workers saying what good practice looks like.
\ldots
[W]ho is leading the conversation around AI and social work?
Not social workers.''}

\subsection{Potential barriers to successful use of AI}

\subsubsection{Lack of automatability}
\looseness=-1
In addition to practitioners' negative experience of current technological systems, another key theme was the sectors' suitability to automation: 
\textit{``I've never felt that social care is built for the information age''} (P8).
In at least four of the interviews, participants raised a direct comparison to the health sector. 
P1---who works across the health and social care divide---explained they are \textit{``more used to these conversations in health care''}.
They shared that \textit{``fantastic things are happening 
\ldots
different skin problems and potential cancers, with AI based systems that can interpret images, and then start to divert people into specialist services.''}
Both accuracy and efficiency were cited as benefits of these applications. 

\looseness=-1
However, P1 went on, \textit{``social care is different from my earlier skin cancer example \ldots [It] is about feeling like you belong in your community, about love and relationships, about living chosen lifestyles''}. 
This sentiment was largely attributed to practitioner behaviours and characteristics. 
For example, \textit{``the behaviours 
\ldots
may have something to do with the way that social work and social care is seen as a non-clinical thing''} (P12).

\vspace{-0.25\baselineskip}
\subsubsection{Lack of funding and infrastructure}

When comparing adoption of AI in social and health care, P1 suggested: 
\textit{``social care is a long way behind partly because it doesn't have the same funding or infrastructure.''}

\vspace{-0.25\baselineskip}
\subsubsection{Challenges around extracting value from data}

\looseness=-1
All participants felt strongly that existing technologies have largely hindered rather than helped practitioners. 
At the heart of these frustrations are ongoing challenges with data:
\textit{``We've got lots of assessments, support plans, letters, complaints, inquiry reports, etc., but it's all in narrative form. 
We can tell you how many reports we've done, but extracting what's actually in them, and how it compares to what the local authority next door is doing, would be really interesting. 
\ldots
[A] question for me is whether that is something machine learning could do''} (P7). 

\looseness=-1
Despite the consensus, there was disagreement around \emph{why} the sector struggles to use data better.
Many of the policy and academic participants felt the fault is with practitioners:
\textit{``The sector is very, very rich in data and information, but they don't understand the value of it, and I think some of that is to do with not needing to. 
\ldots 
If they saw the value of it, I think they would change some of the approaches towards it''} (P12). 
\textit{``There are definitely some quite old school beliefs about the power of intuition. 
\ldots
I think the discourse around data perhaps potentially feels like a challenge to some of those kinds of views''} (P4). 
\textit{``Lots of social workers either don't understand the purpose of it, or why it's helpful. 
They see it as 
\ldots
tick box exercises,
and
the burden of recording 
as
too high. 
I think social workers 
are
cautious of data because 
it's
going to perpetuate inequality,\footnote{P4 referenced the book `Automating Inequality' earlier in the conversation, which warns of the potential of data-driven tools to perpetuate inequality.}
\ldots
[and]
because 
collecting data 
makes it easy
to surveil people''} (P4).	

\looseness=-1
In contrast, some practitioners did not object to data collection per se, but 
did not consider its current uses valuable:  
 %
\textit{``All that we're doing 
is 
\ldots
ticking
the 
boxes.
\ldots
[W]hen our work is authorized, it gets changed. 
It gets formatted differently. 
Words are added so it reads well to our managers. 
\ldots
[P]rofessional judgement doesn't matter. 
I honestly believe it's literally a way for the council to prove that 
\ldots
their social workers are doing what they're told
\ldots
Data collection is not for social workers or for people. 
It's for organizations to show 
\ldots
throughput and growth''} (P8).

\vspace{-0.25\baselineskip}
\section{THEME 2: Where is AI (not) useful}
\label{sec:potential_applications}

The second section of our findings focuses on the perceived opportunities and applications of concern for practitioners.  

\vspace{-0.25\baselineskip}
\subsection{Objections to ADS}

\looseness=-1
As highlighted in \Cref{sec:background}, ADS is increasingly adopted in different jurisdictions and areas of social work.
Despite this widespread adoption in other countries and related domains, all participants felt no desire for AI supported decision making, believing that only humans should be making such high-stakes decisions.

\vspace{-0.25\baselineskip}
\subsubsection{Decisions should be made by humans as a matter of principle}

\looseness=-1
Participants had many thoughts about why ADS should not be used in social work, 
as exemplified by a quote from P11: \textit{``[I]t's important to bear in mind how horrible it can feel to think that decisions about you are being made by a machine, and not by a human.''}

\vspace{-0.25\baselineskip}
\subsubsection{Lack of context, nuance, intuition, and cultural understanding} 

\looseness=-1
Others thought ADS could not handle specific aspects of their decision making.
For example, participants felt ADS does have the intuitions and contextualized understanding of a social worker
\textit{``[W]hen you make an assessment like that, it draws on information which is really difficult to capture,
like the feel of when you go into a house''} (P4),
and would not \textit{``recognise the nuance of a situation''} (P4).
The `nuance' mostly related to the ability to understand the context: \textit{``a tool that could work well in one context might not work well in another''} (P10). 

\looseness=-1
One participant shared an example of how different cultural contexts can impact 
care decisions. 
In the Bangladeshi community, outsourcing care of a family member is often regarded as a family failure,
which an adult social care worker will take into account.
\textit{``[O]ne size doesn't fit all. There's no one approach, no one intervention that you can use that's going to solve all of the many and varied problems that some of my colleagues call the `messiness of human life'.
Each set of individual circumstances is unique''} (P10).

\vspace{-0.25\baselineskip}
\subsubsection{Data does not capture the richness of individual circumstance}

\looseness=-1
Many practitioners view ADS as too reductive:
\textit{``Presumably, any decision-making tool is only as good as the data that goes into it. And---as I mentioned---the data that we're able to gather is dependent on individuals and on subjective judgements. 
\ldots
Something might be lost if the input into the decision was only what you've managed to capture and write down''} (P11). 
%
Doubts about applicability of ADS
in social work due to the reductive nature of data entry were considered in a 2016 report by the Behavioural Insights Team (BIT). 
The report highlighted the large number of factors which could be predictive of potential outcomes, and considered how individual cases are likely to be a unique combination of such factors \citep{tupper2016decision}. 
Still, the BIT concluded that ADS has potential to provide supplementary information to social workers, e.g., by highlighting the likely outcomes of different cases.

\vspace{-0.25\baselineskip}
\subsubsection{Existing ADS target the wrong problems}

\looseness=-1
One participant felt that even if 
the data were not reductive,
ADS would still offer little benefit for high-stakes decisions:
\textit{``[A] tool to help with the big decisions might be solving a problem that doesn't exist, 
because the big decisions are made in collaboration with manager, or in the context of a strategy discussion to consider whether we should have a child protection investigation (they have to make these jointly with health and police), [etc.]
\ldots
So these [big] decisions are already shared, and there are helpful forums for making them.
[In contrast,] the decisions that I find most challenging as a social worker on day-to-day basis are the small [everyday] ones''} (P11). 
This insight further underlines how practitioner requirements
can often be disregarded when considering ADS implementation.

\vspace{-0.25\baselineskip}
\subsubsection{Lack of transparency over current use}
\label{sec:intransparent_use}

\looseness=-1
One participant expressed suspicion that ADS is already being quietly deployed, despite the general resistance amongst our interviewees:
%
\textit{``I actually think that it's happened by stealth \ldots
[N]eeds assessments have an algorithm built into them 
\ldots
that produces [an indicative budget]
based on the answers that the social worker gives as part of the 
assessment. 
The number then forms the basis of the amount of care, monetary-wise, that the person is eligible for, against the Care Act. 
Local authorities have been using that
\ldots
for a number of years, and those algorithms are built into the case management systems.
Now, I don't know where those algorithms come from.
I don't know the monetary aspects to them. 
And I don't know who adds or what weight to what value of whatever you put on''} (P2).

\vspace{-0.25\baselineskip}
\subsection{Areas where AI could potentially add value}

\looseness=-1
When exploring the areas AI-based innovation could benefit social work, a number of recurring use cases emerged.
Importantly, all of the proposals centred around improving the practitioners capacity and ability to provide more meaningful care.
All suggestions sought to either improve upstream processes that were considered outside of their direct roles, or to limit time spent on administrative tasks. 
None of the thirteen participants favoured automating direct care.

\vspace{-0.25\baselineskip}
\subsubsection{Finding information}
\label{sec:finding_info}

\looseness=-1
The most common suggestion, brought up by six participants, was to use AI to find suitable auxiliary support services.
Participants generally wanted to improve the information, access, and relevancy around the initial 
\textit{``provision of information and advice''} (P5).
\textit{``[T]he local authority could do so much better in organizing the information it has about 
all the community assets, and surfacing that information in a way that a member of the public can go into it without even necessarily thinking''} (P3).

\looseness=-1
This type of service would be helpful to both carees and their carers
\textit{``[W]hat I'm thinking about is those situations where, for example, I know that I would like a counselling service that specialises in working with children affected by domestic violence and abuse, who are 11 years old, and boys, and takes referrals from my borough.
These can take a long time to find because it'll be hidden on some random page of a website, or they actually only accept applicants from certain boroughs.
I can definitely see a role for AI there, if that process of going through and looking at all the services could be replaced''} (P11). 

\looseness=-1
Two frontline practitioners described how the existing manual referrals process is limited.
They shared examples of failed ad-hoc attempts to 
improve the process: 
\textit{``I think keeping up to date on local community resources is really hard and time consuming. 
I saw that as an issue in my team and 
I was like: `Why are we each individually googling these resources?'
So I made a spreadsheet to share it. 
But 
I just had to let it go out of date because, realistically, I just couldn't do it''}
(P13). 

\vspace{-0.25\baselineskip}
\subsubsection{Triaging inquiries}

\looseness=-1
The second most discussed use case was AI for triaging inbound enquires,
including both social workers inboxes (P1: \textit{``something that prioritises and flags things to sort out the inbox''}, and inbound service enquiries: 
(P5: \textit{``signpost 
to different services once the system has been able to filter a variety of different information that it had acquired''}). 
The benefit is again reducing the time spent on administrative tasks:
\textit{``[I wonder] whether there're ways of automating some of the signposting and the handoffs, leaving the social worker to focus on the kind of more complex areas of need''} (P1). 

\looseness=-1
However, nearly half of the frontline workers doubted this use case due to accuracy:
\textit{``[A]lthough I do spend a lot of time emailing, I wouldn't say that that's wasted time. 
I think firstly, I'm very happy to be CCed into emails 
\ldots
because everything helps me to be informed about the families that I work with. 
I would struggle to trust that system would work because, for example, I'd say that 70\% of the emails I receive will say at the top something like ‘Be careful, this comes from a sender that you don't normally receive emails from’. 
And the reason is that I just receive emails from so many different agencies.''}
(P11). 

\vspace{-0.25\baselineskip}
\subsubsection{Improving data collection}

\looseness=-1
Another area participants explored was use of technology to improve data collection.
Given we were using AI to transcribe our interviews (\texttt{otter.ai}), we explored if a similar tool would be useful. 
The responses were divided. 
Two practitioners felt strongly about the potential benefits: 
\textit{``If social workers had something like that, you've instantly reduced our workloads by hours and hours a week''} (P8), and \textit{``there's a crying need for data, for evidence, and for some sort of systematic approach to capturing that data on a real life basis''} (P3). 

\looseness=-1
However, two other practitioners had strong reservations about the use of transcription due to the nonverbal signals that they think make up a large part of their interactions: 
\textit{``It's not just the words, it's the feeling that goes behind them.
That's quite difficult to capture''} (P12).
Both also felt the use of transcription would adversely impact the quality of their conversations: 
\textit{``[K]nowing every word is being noted makes things slightly different,
even if it's being turned into a sort of summary rather than a transcription afterwards. 
For example, small casual conversations become less likely, and I think these can sometimes be really important to my overall interaction with a family, as it’s in those sorts of moments when families chat about something that they probably think is completely irrelevant to our work, but it's actually helpful for us to understand''} (P11).

\vspace{-0.25\baselineskip}
\subsubsection{Understanding trends in order to improve service provision}

\looseness=-1
The third most discussed use case centred use of AI for service improvement, and \textit{``resolving some uncertainty''} (P7) through data analysis. 
P7 wanted to 
(a)~\textit{``understand the variation and effects''} in the effectiveness of service provision at the organisational and institutional levels, and
(b)~use AI to identify causes of service provision delays.
These insight could help social workers to allocate resources more appropriately, for example, by adding more shifts at specific times and limiting them at others.
Frontline practitioners working in adult social care 
explained one of their most pressing questions is \textit{``what is it that really makes for a successful move?''} (P7). 
P7 wondered whether AI could use historical data to identify factors which may contribute or hinder an effective move for elderly service users.
For example, the use of machine learning may be able to detect how the existence of a pet impacts the success of a move for service users.

\vspace{-0.25\baselineskip}
\section{THEME 3: Future development}
\label{sec:future_development}

\vspace{-0.25\baselineskip}
\subsection{Participation by practitioners and carees}

\looseness=-1
There was unanimous consensus on the criticality of involving practitioners in the design, development, and deployment of any new technological system. 
All participants felt strongly that \textit{``it all starts from the ground up''} (P8),
and one should \textit{``start by asking people, what is it about the current system you don't like?''} (P8), to ensure people are given \textit{``the chance to actually have their voices heard''} (P8). 
As P12 explained: 
\textit{``It is essential that practitioners are directly involved because they are going to be using it.
You often hear of very clever guys---usually guys, but not necessarily very clever people---developing things and then trying them out, and that's not the way to do it.
There has to be co-production.
You have to engage from the beginning, but not just the practitioners. 
I think you need to engage the families as well, because it's a collaborative process''}. 
The need to involve not only the practitioners was echoed by multiple participants: \textit{``Co-production and ensuring that the lived experience of the service recipient is involved in influencing whatever design or commissioning that's underway''} (P5); 
\textit{``Don't take decisions about us without our engagement''} (P10).

\vspace{-0.25\baselineskip}
\subsection{Overcoming obstacles to participation}

\vspace{-0.25\baselineskip}
\subsubsection{Technological literacy}

\looseness=-1
Many participants felt such participatory design requires first improving data and AI literacy of the involved practitioners:
\textit{``If you don't know what AI is, then you don't know what you don't know. 
You don't know that it's potentially a thing, a tool, a mechanism to proportionately empower or make someone's life better''} (P2). 
For many, this meant more training: \textit{``I think that there should be more in a way of training or literacy around data, or AI, that will help people understand the options.
Help them be more involved in proposing the direction things go, technology-wise''} (P4). 
This digital upskilling should not only be directed at practitioners, but also service users: \textit{``[W]e want to move closer to co-production with the families that we work with. 
But sometimes, a barrier to that is not our access to and familiarity with technology, but the comfort and familiarity with technology of the families''} (P11). 

\vspace{-0.25\baselineskip}
\subsubsection{Champions of practitioner interests}

\looseness=-1
Interviewees also discussed other barriers to adoption of participatory approaches. 
Both practitioner mental energy (P13: \textit{``they just don't have the headspace to take on the challenge''}), and time were mentioned:
\textit{``They don't feel like they've been listened to, so the incentive to volunteer their 
quite precious time
\ldots
is even more reduced. 
For example, we do hear from people who have been involved in things like Clinical Commissioning Groups
\ldots
as the experts by experience, 
and nothing has changed.
\ldots
[They were] there as a token and that's the thing''} (P10).
To overcome these barriers, practitioners 
suggested creating champions who could be upskilled to understand more about AI and technology:
\textit{``[E]ngage a small group of people who are really interested, and allow them to be the voice of social workers.''} (P11).

\vspace{-0.25\baselineskip}
\subsubsection{Compensating practitioners}

\looseness=-1
To address the time commitment issue, P10 suggested:
\textit{``[T]here's an argument to say if people are giving up their time like this, then they should could be recompensed for that in some way''}.

\vspace{-0.25\baselineskip}
\subsection{Design requirements}

\vspace{-0.25\baselineskip}
\subsubsection{Agreed objectives}

\looseness=-1
Having clear agreed objectives was noted as a precodition to practitioner engagement:
\textit{``Developers, commissioners, practitioners---they all have different end goals 
\ldots
[F]ind some kind of space where they understand each other and the goals overlap''} (P10). 
\textit{``[B]e clear about what you're trying to achieve 
\ldots
and try to agree what success would look like, and why new and emerging technology is a good way of trying to deliver these things''} (P1). 

\looseness=-1
P1 also
shared an example of the impact when this approach is not adopted:
 %
\textit{``Everyone had gone into 
[the project]
hoping it would produce something slightly different for them. 
And everyone ended up disappointed because no one way of doing things could ever achieve all those different aims. 
Most people had an implicit assumption about what [the technology] was supposed to achieve. 
But they hadn't really articulated it to themselves, let alone to anyone else''} (P1). 
 
\vspace{-0.25\baselineskip}
\subsubsection{Adding value and reducing burden} 

\looseness=-1
Existing tools are often considered burdensome (\Cref{sec:experience}), failing to meet practitioner expectations.
P4 suggested: 
\textit{``I think it should have started from
\ldots
what needs to be recorded.
A system was designed 
\ldots
[as a] workflow thing,
whereas it should just be a minimum of what you need 
\ldots
[and]
make it as easy as possible''}.

\vspace{-0.25\baselineskip}
\section{Discussion and implications}
\label{sec:discussion}

Three key topics emerged from our interviews with the UK social work practitioners:
\begin{enumerate}[(i),topsep=0pt]
    \item \looseness=-1
    Largely negative past experiences with technological systems incorporated into social work. 

    \item \looseness=-1
    Unanimous aversion to the incorporation of ADS systems.

    \item \looseness=-1
    An interest in non-ADS AI applications, conditioned on these being developed in a participatory manner. 
\end{enumerate}
In this section, we explore these key findings further and situate them within the wider literature (\S\ref{sec:key_find_lit}), discuss visions for future technology design in social work (\S\ref{sec:design_visions}), and important considerations for future developments (\S\ref{sec:considerations}).

\vspace{-0.25\baselineskip}
\subsection{Exploring the key findings} 
\label{sec:key_find_lit}

\vspace{-0.25\baselineskip}
\paragraph{Little benefit to practitioners.} \looseness=-1
Our interviewees perceive technology as rarely helpful to social workers. 
Existing literature similarly found attempts to digitise and automate
\textit{``highly problematic for frontline social workers, particularly in terms of diverting their time, attention, and energy away from direct work with service users''} \citep{munro2011munro,white2009whither,gillingham2021practitioner}. 
Ongoing challenges with data collection were flagged as a core frustration, agreeing with \citet{gillingham2021practitioner} who found data capture and entry are perceived as a tax and a form of surveillance.
The amount of time that practitioners spend manually repeating data entry has become disproportionate to the time spent giving care. 
The data collected is also not perceived as useful, rendering the effort meaningless.
Similar sentiment extends beyond social work, with frustration over electronic record keeping affecting probation officers \citep{mair2006worst} and doctors \citep{khairat2020association, gawande2018doctors} alike.
While the user experience of data collection improved in the commercial world, creating adaptive and smooth-looking forms, the systems used by social care practitioners stayed painfully behind. 

\vspace{-0.25\baselineskip}
\paragraph{Lack of practitioner-centred design} \looseness=-1
Involving users and impacted families in the design process is considered critical to success of new digital technologies \citep{wagner2007moving,delgado2021stakeholder}. 
Yet our interviewees detailed numerous failures
largely due to misalignment with the users' wants, needs, or available resources. 
The typical cause is designers' lack of understanding of frontline social work practice, combined with insufficient involvement of social workers in the design \citep{wagner2007moving}.
Similar to \citep{kawakami2022care}, our findings suggest relevant stakeholders need to be involved not only in interface design, but in establishing legitimacy of the innovation itself.
Moreover, as our interviewees highlighted, practitioners are not the only stakeholders that should be involved in the design process.
\citet{scott2022algorithmic} ran two workshops to identify the needs of migrant job seekers, so that these could inform the design of Public Employment Services algorithmic systems. 
Similarly,  
\citet{cowan2021rapid} ran a multi-stakeholder workshop to prioritise innovations for adult social care. 
However, they found that it difficult to resolve tensions between different stakeholders in a single workshop;
practitioners may
also
not speak freely around management.

\vspace{-0.25\baselineskip}
\paragraph{Negative experiences shape practitioners' sentiment towards future innovation} \looseness=-1
As in \citep{brown2019toward,gillingham2015electronic}, our interviewees generally distrust existing digital tools in their workplace, which contributes to their suspicion of future technological innovation. 
In contrast to the long-standing misconception that social workers' `computer phobia' is a key cause of the failure of digital technologies in the sector \citep{neugeboren1996organizational}, our findings indicate that frontline practitioners are happy to embrace new technology in their private lives, and do not have an inherent aversion to innovation. 
Instead, negative past experiences have contributed to widespread suspicion of overhyped technological claims, and deep-rooted scepticism around social work's compatibility with digital technology.
Similar notions have been proposed by \citet{gillingham2015electronic} almost a decade ago;
the distance between the quality of personal and work technology seems to have only grown starker. 

\vspace{-0.25\baselineskip}
\paragraph{Opposition to ADS} \looseness=-1
As outlined in \Cref{sec:background}, adoption of ADS within social work and welfare is increasing worldwide \citep{saxena2020human,sleep2022adm}.
It is therefore significant that our findings reveal no desire for ADS among practitioners. 
While past work found dissatisfaction with specific tools \cite{kawakami2022care,kawakami2022improving}, we find interviewees actually oppose ADS in general.
Strikingly, the agreement on this point was unanimous;
while our sample may not be fully representative,
we have no reason to suspect it was biased in this specific direction.
%
The reasoning behind these objections is clear:
practitioners felt their work requires contextualisation and nuance unachievable by ADS. 
In line with \citep{brown2019toward,kawakami2022improving}, practitioners thought it impossible
for a model to adequately account for all relevant decision elements, explicit and tacit,
even if some research suggests ADS has the ability to assist social workers' decision making (e.g., by surfacing relevant information from other cases) \citep{dare2017ethical}.

\vspace{-0.25\baselineskip}
\subsection{Design visions}
\label{sec:design_visions}
\looseness=-1
The design visions of our interviewees all relate to desire to provide better care.
Several suggestions focused on improving auxiliary support which they felt beyond their ability or capacity to provide.
This included better infrastructure for community and peer support, technology to help route direct care requests, and improved access to information regarding available services. 
As for assistance for practitioners, most wished for technologies that would allow them to dedicate more time to their carees.
%
Here, we discuss the most popular suggestions in more detail.
We emphasize that we wanted to give voice to all our interviewees, disregards of their level of experience with AI; the interview was designed to provide adequate support.
While social workers often know their clients well, we stress that the carees and others impacted should be involved in design process. 
Such consultations present critical future work.

\vspace{-0.25\baselineskip}
\paragraph{Directories and recommender systems for ancillary support.}
\looseness=-1
Practitioners stressed they often do not have sufficient time to help individuals find personalised auxiliary support. 
Recommender systems are used in various settings to give users personalised recommendations based on their profile and preferences. 
It was proposed (\S\ref{sec:finding_info}) that directories or recommender systems could be integrated into existing social care platforms, websites, or chatbots to provide more personalised information on access to support.
An automated `matching' service for accessing peer support was also envisioned. 
The primary purpose of these visions is to improve the quality and efficiency of upstream care.
A secondary benefit is a time saving for social workers themselves,
who often manage very large caseloads.

\vspace{-0.25\baselineskip}
\paragraph{Re-envisioning data capture and entry}
\looseness=-1
Frontline practitioners wanted tools that would help lift the burden of administrative tasks such as transcription and data entry, without trying to influence their decision-making.
Currently, practitioners feel data input is not designed in a user-centric way, and they are unable to collect, record, and flag important information.
More flexible data collection, where practitioners have more agency 
could be considered. 
One suggestion was to use of natural language processing to semi–automatically populate forms and thus help practitioners cut back on their administrative time. 
Instead of manually inputting the same information multiple times into different systems or forms, 
the system would populate duplicate fields.
Practitioners could then review this information and edit where appropriate. 

\vspace{-0.25\baselineskip}
\paragraph{Improving practitioner agency and data quality.}
\looseness=-1
Our findings reveal how current `check-list' decision-making, such as those involved in adult social care (\S\ref{sec:intransparent_use}), do not align with practitioner needs.
As a result, practitioners may try and `game the system' by inputting the information they know will provide the output they desired, even if it does not reflect reality. 
This, in turn, influences data quality and further masks the ways the current system does not cater to certain needs.
To overcome this, one design vision suggested reversing the data capture process. 
The first step will be for practitioners to enter the required budget, and the second step enter justification. 
The reversal can be implemented only as an option, for practitioners disagree with the suggested budget.
The data can then be used to update the budget-setting model.

\vspace{-0.25\baselineskip}
\subsection{Considerations for future developments}
\label{sec:considerations}

Alongside the design visions, we make the following suggestions based on the interviews:
\begin{itemize}[leftmargin=5.5mm,topsep=0.5pt]
    \item \looseness=-1
    \textbf{Earn and build practitioner trust}: Understand historic experience and failings in technological deployments, take steps to repair trust, and ensure prior mistakes are not repeated.
    
    \item \textbf{Prioritise participation}: Prioritise participation from practitioners and those they serve, and overcome obstacles to participation through training, compensation, and other incentives.
    
    \item \looseness=-1
    \textbf{Be participant-centred}:  
    (i)~Agree on objectives, 
    (ii)~focus on user experience and low burden of use, 
    (iii)~carefully consider what data should be captured and how, and 
    (iv)~ensure the tool brings value to practitioners and carees.
    
    \item \looseness=-1
    \textbf{Consider where AI should (not) be applied}:
    Instead of ADS, consider applications that can help social workers with administrative burden,
    understanding resource demands, and factors that lead to successful interventions. 
    Ask participants which tasks are (not) amenable to technological support. 
    Discussions about where AI should (not) be applied should centre those impacted by the systems, particularly those in the care system and their families.
    
    \item 
    \textbf{Tackle barriers to success}: Address wider barriers including improved funding, infrastructure, and data literacy. 
    Consider methods that could extract value from the large amounts of unstructured data available.
    First and foremost, discuss with practitioners how potential tools could bring value to them and those they serve.
\end{itemize}

Finally, while social workers can have deep insight into the needs of their clients, it is critical carees and their families are consulted independently when considering future developments.

\vspace{-0.25\baselineskip}
\section{Conclusion}
\label{sec:conclusion}

\looseness=-1
As the integration of AI technologies into the UK's social care sector continues to gain momentum, it is imperative that we look beyond the developer promises,
and consider the opinions of those impacted by the introduction of these technologies.
We conducted thirteen semi-structured interviews, with the goal of understanding the experiences and attitudes to AI among UK social care professionals, and how these shape their views on the development of future systems.

\looseness=-1
We found practitioners have been largely dissatisfied with technology deployed in social care, particularly with algorithmic decision systems, and the burden introduced by data collection requirements.
Many of these failures were caused by a lack of involvement of practitioners in the design process, resulting in applications that solve no problem relevant to the carers or carees.
While these past experiences engendered general scepticism, there is still significant interest in AI tools that could reduce administrative burdens, and allow practitioners to spend more time helping people in need. 

\looseness=-1
We made several recommendations based on our findings. 
First, we advocate for a strong commitment to participatory design, ensuring that the perspectives and needs of practitioners and carees are incorporated throughout the development and deployment.
Second, we recommend a refocus towards applications genuinely supporting and complementing social care professionals, rather than replacing or undermining their expertise.
Lastly, we highlight the importance of transparency, and of addressing wider barriers to successful use of technology, including funding, infrastructure, and data literacy.
Our conclusions were shared with the practitioners and BASW, and we hope they will lead to development of more effective and safe AI systems, helping social care professionals in their unwavering quest to support vulnerable individuals and communities. 



\bibliographystyle{ACM-Reference-Format}
\bibliography{bib}

\newpage
\appendix
\beginsupplement
\onecolumn


\includepdf[scale=0.8,pages=1,pagecommand=\section{Participant information sheet (redacted)}\label{apx:info_sheet}]{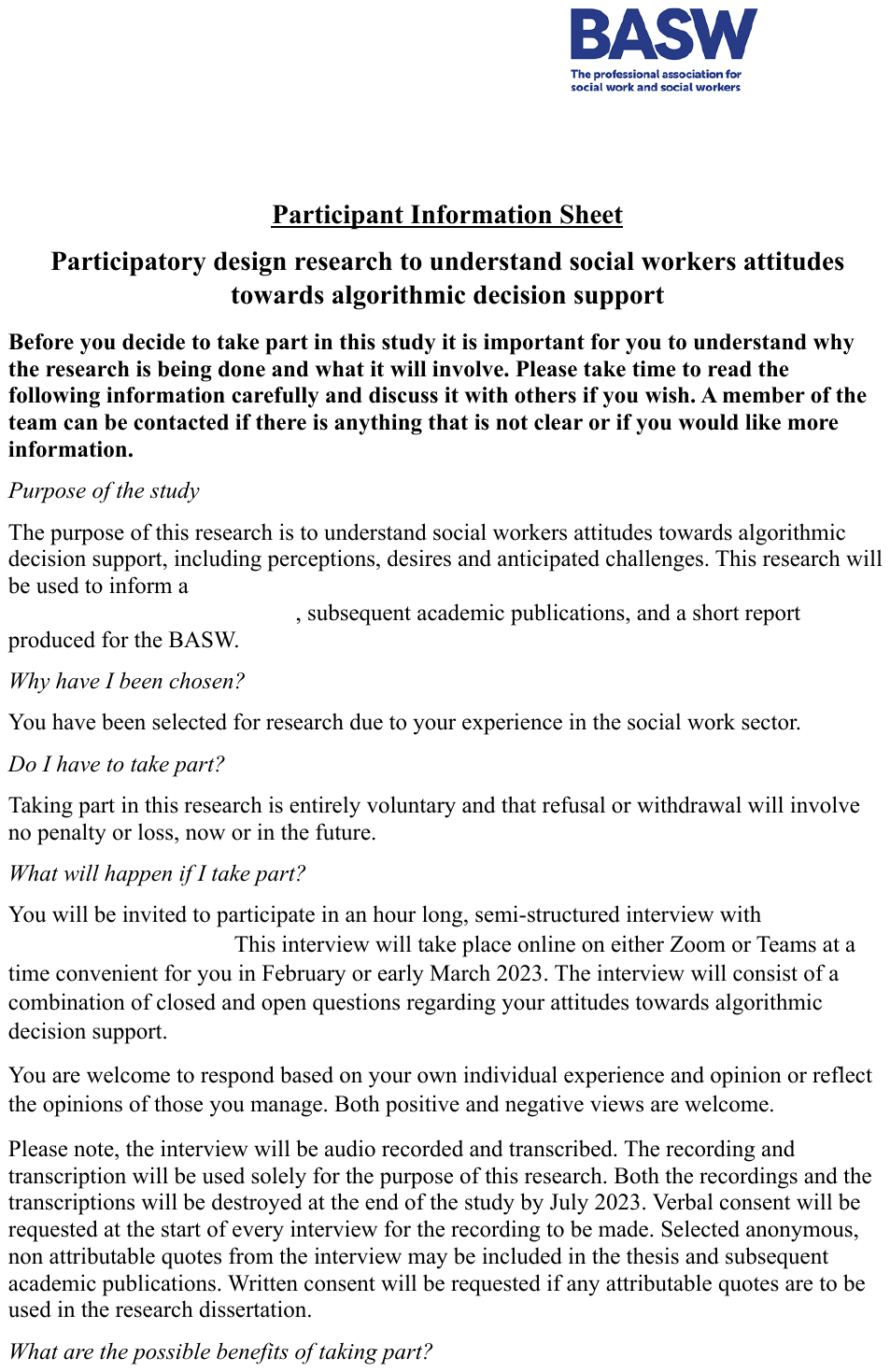}
\includepdf[scale=0.8,pages=2-]{Participant-information-sheet_Redacted.pdf}


\includepdf[scale=0.8,pages=1,pagecommand=\section{Interview guide}\label{apx:guide}]{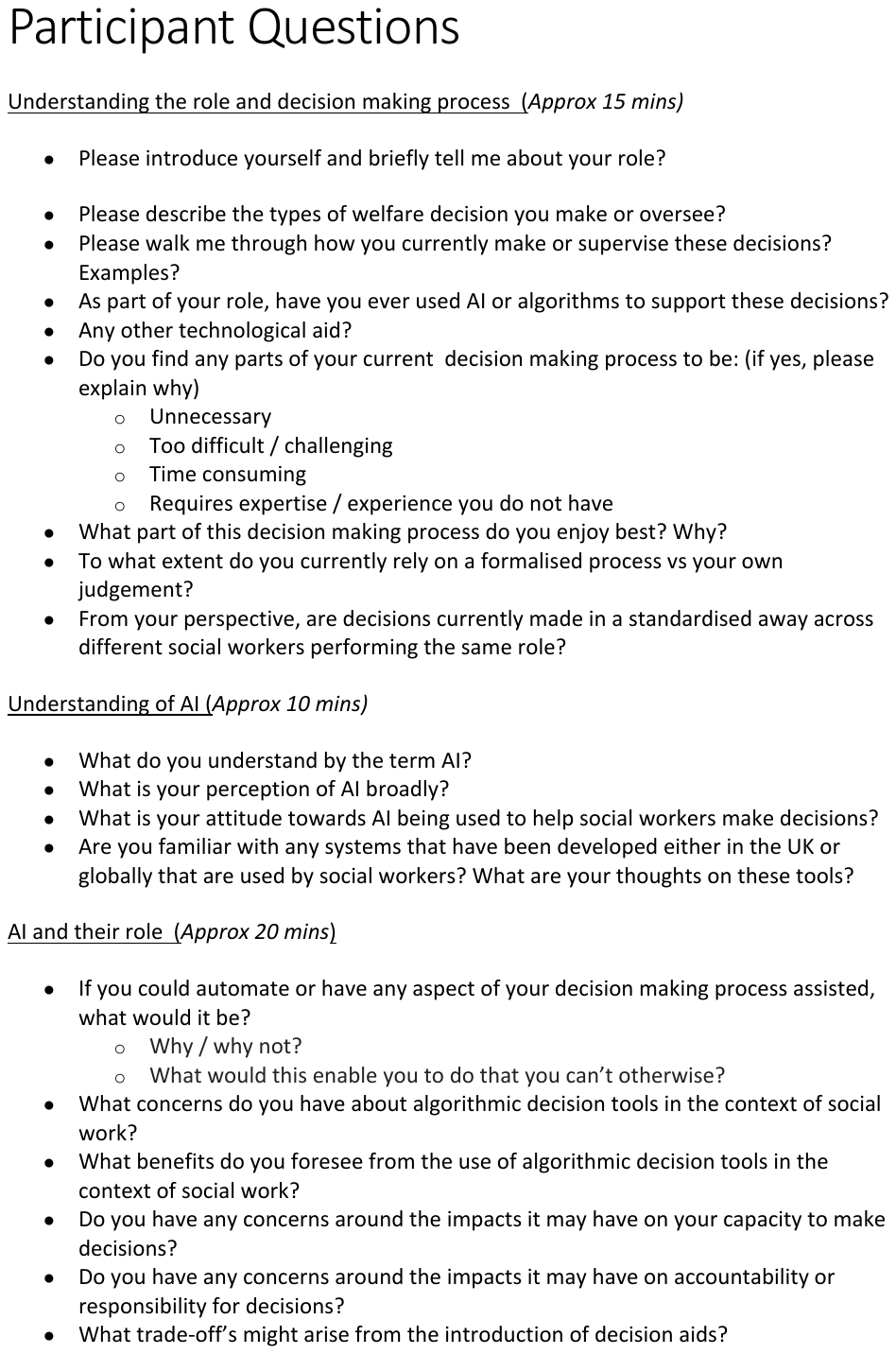}
\includepdf[scale=0.8,pages=2-]{Participant_semi-structured_interview_questions.pdf}

\end{document}